\documentstyle{article}
\begin{document}

\baselineskip=23pt

\begin{flushleft}
{\bf {\Huge 
Formulas for direct determination of real speed in Galactic superluminal
sources
}}\\
\vspace{2mm}

{\bf Yi-Ping Qin$^{1,2,3,4}$ }

{\bf $^{1}$ Yunnan Observatory, Chinese Academy of Sciences, Kunming, Yunnan
650011, P. R. China; E-mail: ypqin@public.km.yn.cn }

{\bf $^{2}$ National Astronomical Observatories, Chinese Academy of Sciences 
}

{\bf $^{3}$ Chinese Academy of Science-Peking University joint Beijing
Astrophysical Center }

{\bf $^{4}$ Yunnan Astrophysics Center } \\
\end{flushleft}
\vspace{6mm}

In this paper we show explicit and direct relations between the expected
quantities and the observed quantities for the components of Galactic
superluminal sources. Basic formulas for calculating the real speed and the
angle to the line of sight of the components from the data of proper motions
and the distance of the source are presented. We point out that the real
speed and the angle to the line of sight of components can be uniquely and
directly determined from the observed values of the proper motions and the
distance of the source. It is not necessary to calculate an intermediate
quantity first and then using the resulted value to calculate the two
required quantities. The process of the calculation is simple, and in this
way, some extra uncertainties are avoided.

\begin{flushleft}
\end{flushleft}

\section{Introduction}

Superluminal motions found in distant quasars and active galactic nuclei
(AGNs) are an interesting astrophysical phenomenon that arouses a worldwide
attention. The motions have been inferred for radio emitting components in
these objects (e.g., Blandford et al., 1977). The components move away from
the central sources at apparent speeds greater than that of light $c$. A
generally accepted explanation is that clouds of plasma are ejected in
opposite directions from the central source at real speeds close to $c$, and
the relativistic effects lead to the apparent superluminal motion (Rees,
1966). But the extreme distance of the sources introduces many uncertainties
into this interpretation (Kellermann and Owen, 1988). Fortunately, there is
now direct evidence for superluminal motions in the radio images of two
strong Galactic X-ray transient sources, GRS 1915+105 and GRO J1655-40
(Mirabel and Rodriguez, 1994; Tingay et al., 1995; Hjellming and Rupen,
1995). As pointed out by Mirabel and Rodriguez (1994), the optical, infrared
and X-ray properties of GRS 1915+105 (Mirabel et al., 1984; Harmon et al.,
1994) suggest that the source is either a neutron star or a black hole that
is ejecting matter in a process similar to, but on a smaller scale than that
seen in quasars. Because of their relative proximity, these Galactic
superluminal sources may offer the best opportunity to gain a general
understanding of relativistic ejections seen elsewhere in the Universe.

To understand the power of the sources that are ejecting material in such a
violent manner, the measure of the real speed of the components is desired.
The difficulty of measuring the real speed of components of quasars and AGNs
is that distances to these sources are still uncertain since they depend on
the cosmological parameters and on the cosmology itself. Since the distance
of Galactic sources is free of cosmological models and can be directly
measured, the discovery of superluminal motions within our own Galaxy makes
it possible to determine the real speed of the components at a high level of
accuracy.

In the following we illustrate how to use the measure of the proper motions
of components to determine their real speeds as well as the angles to the
line of sight.

\section{Determination of the real speed of components of Galactic sources}

For Galactic sources, the apparent proper motions of the approaching and
receding components moving in opposite directions away from a common origin
of a source along an axis at an angle $\theta $ to the line of sight at the
real speed $\beta c$ can be expressed as (Pearson and Zensus, 1987; Mirabel
and Rodriguez, 1994) 
\begin{equation}
\mu _a=\frac{\beta \sin \theta }{1-\beta \cos \theta }\frac cD,
\end{equation}
\begin{equation}
\mu _r=\frac{\beta \sin \theta }{1+\beta \cos \theta }\frac cD,
\end{equation}
where $D$ is the distance from the observer to the source, and $\mu _a$ and $%
\mu _r$ are the proper motions of the approaching and receding components,
respectively. These two equations lead to 
\begin{equation}
\beta =\frac{\sqrt{\left( \frac{2\mu _a\mu _rD}c\right) ^2+(\mu _a-\mu _r)^2}
}{\mu _a+\mu _r},
\end{equation}
\begin{equation}
\theta =\arccos \frac{\mu _a-\mu _r}{\sqrt{\left( \frac{2\mu _a\mu _rD}%
c\right) ^2+(\mu _a-\mu _r)^2}}.
\end{equation}
It is clear that, when the proper motions $\mu _a$ and $\mu _r$ as well as
the distance $D$ are measured, the real speed $\beta c$ and the angle to the
line of sight $\theta $ can then be uniquely determined.

In practice, the observed parameters $\mu _a$, $\mu _r$ and $D$ are always
presented in the form $\mu _a\pm \triangle \mu _a$, $\mu _r\pm \triangle \mu
_r$ and $D\pm \triangle D$. If the errors are relatively small, the errors
of $\beta $ and $\theta $ can be determined by means of: 
\begin{equation}
\triangle \beta =\frac{\sqrt{\left[ \frac{4D\mu _a^2\mu _r^2(\mu _a+\mu _r)}{%
c^2}\right] ^2\triangle D^2+\left[ \frac{4D^2\mu _a\mu _r^3}{c^2}+2\mu
_r(\mu _a-\mu _r)\right] ^2\triangle \mu _a^2+\left[ \frac{4D^2\mu _a^3\mu
_r }{c^2}-2\mu _a(\mu _a-\mu _r)\right] ^2\triangle \mu _r^2}}{\beta (\mu
_a+\mu _r)^3},
\end{equation}
\begin{equation}
\triangle \theta =\frac{\frac{2D\mu _a\mu _r}c\sqrt{(\mu _a-\mu _r)^2\left( 
\frac{\triangle D}D\right) ^2+\mu _r^2\left( \frac{\triangle \mu _a}{\mu _a}%
\right) ^2+\mu _a^2\left( \frac{\triangle \mu _r}{\mu _r}\right) ^2}}{\frac{%
4D^2\mu _a^2\mu _r^2}{c^2}+(\mu _a-\mu _r)^2}.
\end{equation}

Recently, two Galactic sources were found to have bidirectional relativistic
proper motions of radio components and the proper motions as well as the
distances of the sources were well measured. They are: (1) GRS 1915+105, $%
D=(12.5\pm 1.5)kpc$, $\mu _a=(17.6\pm 0.4)masd^{-1}$, and $\mu _r=(9.0\pm
0.1)masd^{-1}$ (Mirabel and Rodriguez, 1994); (2) GRO J1655-40, $D=3.2kpc$, $%
\mu _a=54masd^{-1}$, and $\mu _r=45masd^{-1}$ (Hjellming and Rupen, 1995).
With these data, the values of the real speed and the angle to the line of
sight for these sources can then be uniquely determined: (1) GRS 1915+105, $%
\beta c=(0.917\pm 0.098)c$, $\theta $$=(69.36\pm 2.35)^{\circ }$; (2) GRO
J1655-40, $\beta c=0.910c$, $\theta $$=84.27^{\circ }$. The errors of $\beta 
$ and $\theta $ for GRO J1655-40 are not given since $\triangle \mu _a$, $%
\triangle \mu _r$ and $\triangle D$ for the source are not available.

The above method can be applied to extragalactic sources when $D$ is
replaced by $D_L/(1+z)$, where $D_L$ and $z$ are the luminosity distance and
the redshift of the source, respectively. In the Friedmann cosmology, $D_L$
is determined by 
\begin{equation}
D_L=\frac{cz(1+z+\sqrt{1+2q_0z})}{H_0(1+q_0z+\sqrt{1+2q_0z})},
\end{equation}
where $H_0$ and $q_0$ are the Hubble constant and the deceleration parameter
of the universe, respectively. In this case, 
\begin{equation}
D=\frac{cz(1+z+\sqrt{1+2q_0z})}{H_0(1+z)(1+q_0z+\sqrt{1+2q_0z})}.
\end{equation}
To apply Equations (5) and (6), one just replaces the observed $\triangle D$
with a calculated one, derived from Equation (8). That is 
\begin{equation}
\triangle D=\sqrt{(\frac{\partial D}{\partial H_0})^2\triangle H_0^2+(\frac{%
\partial D}{\partial q_0})^2\triangle q_0^2+(\frac{\partial D}{\partial z}%
)^2\triangle z^2},
\end{equation}
with 
\begin{equation}
\frac{\partial D}{\partial H_0}=-\frac D{H_0},
\end{equation}
\begin{equation}
\frac{\partial D}{\partial q_0}=-\frac{Dz[1+z+q_0z+(1+z)\sqrt{1+2q_0z}]}{%
(1+2q_0z)(2+z+q_0z)+(2+z+3q_0z+q_0z^2)\sqrt{1+2q_0z}},
\end{equation}
\begin{equation}
\frac{\partial D}{\partial z}=\frac{%
D[2+2z+z^2+4q_0z+2q_0z^2+q_0^2z^2+q_0z^3-q_0^2z^3+(2+2z+z^2+2q_0z)\sqrt{%
1+2q_0z}]}{z(1+z)[(1+2q_0z)(2+z+q_0z)+(2+z+3q_0z+q_0z^2)\sqrt{1+2q_0z}]},
\end{equation}
where $\triangle H_0$, $\triangle q_0$ and $\triangle z$ are the errors of $%
H_0$, $q_0$ and $z$, respectively.

\section{Discussion and conclusions}

In this paper we show that the real speed and the angle to the line of sight
of components of Galactic sources can be uniquely and directly determined
from the measured values of the proper motions and the distance of the
sources. It is not necessary to calculate an intermediate quantity, e.g. $%
\beta \cos \theta $, first and then using the resulted value to calculate
the two required quantities. If so, not only the process of calculation is
more complicated but also an extra uncertainty, coming from the intermediate
step of calculation, might be brought into the final calculation and hence
might affect the final result. In the paper of Mirabel and Rodriguez (1994),
they used the same data to calculate the quantity $\beta \cos \theta $ first
and then used this value to calculate $\beta $ and $\theta $. Their
calculation gave $\beta \cos \theta =0.323\pm 0.016$, $\beta =0.92\pm 0.08$,
and $\theta $$=(70\pm 2)^{\circ }$. There are slight differences between
theirs and ours. The differences might come from the extra uncertainty of $%
\beta \cos \theta $ resulted from the intermediate step of calculation, or
might come from other aspects. The value of $\triangle \beta $ derived in
this paper is about $20\%$ larger than that from Mirabel and Rodriguez
(1994). We notice that $\triangle D$ is slightly larger than $10\%$ of $D$
for the source ($12\%$). This might be the cause for the difference. In this
kind of case, the method of Mirabel and Rodriguez (1994) might probably be
better.

This paper presents explicit and direct relations between the desired
quantities $\beta $, $\theta $ and the measurements of $\mu _a$, $\mu _r$
and $D$. In this way, not only the process of calculation is simple, but
also some extra uncertainties are avoided.

\vspace{20mm}

{\bf ACKNOWLEDGEMENTS}

It is my pleasure to thank G. Z. Xie, Xue-Tang Zheng and Shi-Min Wu for
their guide and help. This work was supported by the United Laboratory of
Optical Astronomy, CAS, the Natural Science Foundation of China, and the
Natural Science Foundation of Yunnan.

\newpage

\begin{verse}
{\bf REFERENCES}

Blandford, R. D., McKee, C. F., and Rees, M. J.: 1977, {\it Nature} {\bf 267}%
, 211.

Harmon, A. et al.: 1994, in: C. E. Fichtel, N. Gehrels and J. P. Norris
(eds.), {\it AIP Conf. Proc. No. 304}, Am. Inst. of Physics, New York, p.
210.

Hjellming, R. M., and Rupen, M. P.: 1995, {\it Nature} {\bf 375}, 464.

Kellermann, K. I., and Owen, F. N.: 1988, in: G. L. Verschuur and K. I.
Kellermann (eds.), {\it Galactic and Extragalactic Radio Astronomy},
Springer, New York, p. 563.

Mirabel, I. F., et al.: 1984, {\it Astr. Astrophys.} {\bf 282}, L17.

Mirabel, I. F., and Rodriguez, L. F.: 1994, {\it Nature} {\bf 371}, 46.

Pearson, T. J., and Zensus, J. A.: 1987, in: J. A. Zensus and T. J. Pearson
(eds.), {\it Superluminal Radio Sources}, Cambridge Univ. Press, Cambridge,
p. 1.

Rees, M. J.: 1966, {\it Nature} {\bf 211}, 468.

Tingay, S. J., Jauncey, D. L., and Preston, R. A. et al.: 1995, {\it Nature} 
{\bf 374}, 141.
\end{verse}

\end{document}